\documentclass[aps,pra,twocolumn,superscriptaddress,showpacs,floatfix,amsmath,amssymb]{revtex4-2}
\usepackage[breaklinks=true,colorlinks,citecolor=blue,linkcolor=blue,urlcolor=blue]{hyperref}
\usepackage{graphicx}
\usepackage{color}
\usepackage{braket}
\usepackage{siunitx}
\usepackage{bm}
\usepackage{makecell}
\usepackage[normalem]{ulem}
\usepackage[usenames,dvipsnames]{xcolor}

\newcommand*{\citen}[1]{%
  \begingroup
    \romannumeral-`\x 
    \setcitestyle{numbers}%
    \cite{#1}%
  \endgroup
}

\begin{document}

\title{Using gradient-based algorithms to determine ground
state energies \\ on a quantum computer}

\author{T. Piskor}
\email{tomislav.piskor@quantumsimulations.de}
\affiliation{HQS Quantum Simulations GmbH, Haid-und-Neu-Strasse 7, 76131 Karlsruhe, Germany}
\affiliation{Theoretical Physics, Saarland University, 66123 Saarbr\"ucken, Germany}

\author{F. G. Eich}
\affiliation{HQS Quantum Simulations GmbH, Haid-und-Neu-Strasse 7, 76131 Karlsruhe, Germany}

\author{J.-M. Reiner}
\affiliation{HQS Quantum Simulations GmbH, Haid-und-Neu-Strasse 7, 76131 Karlsruhe, Germany}

\author{S. Zanker}
\affiliation{HQS Quantum Simulations GmbH, Haid-und-Neu-Strasse 7, 76131 Karlsruhe, Germany}

\author{N. Vogt}
\affiliation{HQS Quantum Simulations GmbH, Haid-und-Neu-Strasse 7, 76131 Karlsruhe, Germany}

\author{M. Marthaler}
\affiliation{HQS Quantum Simulations GmbH, Haid-und-Neu-Strasse 7, 76131 Karlsruhe, Germany}

\author{F. K. Wilhelm}
\affiliation{Theoretical Physics, Saarland University, 66123 Saarbr\"ucken, Germany}

\begin{abstract}
  Variational algorithms are promising candidates to be implemented on near-term quantum computers. The variational quantum eigensolver (VQE) is a prominent example, where a parametrized trial state of the quantum mechanical wave function is optimized to obtain the ground state energy. In our work, we investigate the variational Hamiltonian Ansatz (VHA), where the trial state is given by a non-interacting reference state modified by unitary rotations using generators that are part of the Hamiltonian describing the system. The lowest energy is obtained by optimizing the angles of those unitary rotations. A standard procedure to optimize the variational parameters is to use gradient-based algorithms. However, shot noise and the intrinsic noise of the quantum device affect the evaluation of the required gradients. We studied how different methods for obtaining the gradient, specifically the finite-difference and the parameter-shift rule, are affected by shot noise and noise of the quantum computer. To this end, we simulated a simple quantum circuit, as well as the 2-site and 6-site Hubbard model.
\end{abstract}

\maketitle

\section{Introduction}
Recent advances in quantum computing hardware \cite{arute2019,zhong2020,pino2020} have led to an increasing interest in developing quantum algorithms for noisy intermediate scale quantum (NISQ) computers.
One of the most promising applications for these near-term quantum computers is the simulation of fermionic systems, such as lattice systems or small molecules. The determination of the ground state and ground-state energy is one of the key aspects to gain information about the system. Hybrid quantum-classical algorithms, such as the Variational Quantum Eigensolver (VQE) \cite{peruzzo2014} turn out to have low resource requirements making them suitable for NISQ computers. These algorithms determine the ground state energy of a system via preparing a parametrized trial state on a quantum computer and optimizing the parameter set with a classical optimization routine.

The definition of the trial state depends on the choice of the VQE Ansatz. One Ansatz that is well suited for lattice systems is the Variational Hamiltonian Ansatz (VHA) \cite{wecker2015}, stating that for each term of a generic decomposition of the Hamiltonian a separate parameter is defined. This way, the number of parameters, and thus the circuit depth can be reduced compared to other VQE Ans\"atze, such as the unitary coupled cluster with single and double excitations (uCCSD) \cite{romero2018}. In the uCCSD Ansatz parameters are defined based on single and double excitations of electrons from occupied to unoccupied (molecular) orbitals thus the number of parameters scales polynomially with the system size -- with the excitation level determining the highest power of the polynomial scaling. By contrast, the VHA uses the structure of the Hamiltonian to determine the variational form. Therefore periodic lattice Hamiltonians, which are typically characterized by a sparse Hamiltonian with few independent coupling constants yield an Ansatz which does not scale with the system size.

The VHA is inspired by an adaptation of the so-called ``adiabatic connection'' from many-body perturbation theory \cite{FetterValecka:71,GiulianiVignaleLRPT:05}. Starting from the solution (ground state) of non-interacting electrons the state is evolved on the quantum computer using a sequence of unitary propagators constructed using parts of the fully-interacting Hamiltonian. The variational parameters correspond to the propagation times of each unitary operator. Having opted for the VHA to provide the parametrized trial state, the parameter set is optimized such that the expectation value of the interacting Hamiltonian is minimized. There are several approaches to determine the optimal parameter set for a given cost function ranging from gradient-free \cite{powell2007,powell1994} to gradient-based algorithms \cite{nocedal2006}, which usually lead to faster convergence compared to the gradient-free alternatives. In the present work we focus on the analysis of the evaluation of gradients on NISQ computers, which means that we are less concerned about the specific gradient-based optimization algorithm. Hence, we use a simple steepest-descent approach for the optimization, where the parameters are updated using the gradient of the energy directly, employing a fixed learning rate (damping of the gradient).

The main question we address in this work is how measurement statistics and noise affect the optimization of the parameters appearing in the quantum circuit. Since we always have to perform measurements in order to get the expectation value of an observable, shot noise will always be existent -- even on error-corrected quantum devices. However, when simulating the quantum device we can obtain the limit of an infinite number of measurements, providing us with the ideal reference result for the algorithm. Furthermore, due to adverse coupling of the qubits to the environment, another source of error has to be considered, namely the intrinsic depolarization of qubits. However, we find that the dominant error for the noise model chosen in our simulations stems from measurement statistics.

For the determination of the cost function's gradient we compare two procedures: 1) the finite-difference approximation to compute the gradients, which we also refer to as ``numerical'' method, 2) the so-called parameter-shift rule, which we also refer to as ``analytical'' method \cite{schuld2019,mari2020,crooks2019,banchi2020,meyer2020}. While calculating the gradient numerically, the outcome of the result is more susceptible towards noise effects, such as shot noise or depolarization effects occurring in qubits, because there is a competition of improving the numerical gradient by reducing the finite step size to evaluate the gradient versus resolving differences of the cost function evaluated for two nearby values of a given parameter. To bypass this hurdle, gradients can also be determined analytically via the parameter-shift rule, with the hope to show a more resilient behavior \cite{schuld2019} to noise effects, because it relies on a \emph{fixed} finite difference. However, there is an important caveat for using the parameter-shift rule for optimizing a VHA: Compared to the numerical case, where the number of additional circuit evaluations only scales with the number of parameters, the number of additional circuit evaluations scales with twice the number of parametrized gates for the parameter-shift rule. While the number of parameters solely depends on the Ansatz, the number of parametrized gates depends on the implementation of the trial state in the quantum circuit. In general it has to be considered that one parameter might occur in multiple parametrized gates -- especially in case of the VHA (see section \ref{sct:vha}) -- and thus the number of parameters might be much smaller than the number of parametrized gates.

In the following we compare both methods for determining the gradients, using a simple gradient-based optimization algorithm, by investigating a simple one-qubit circuit and a 2- and 6-site Hubbard model, mapped onto 4 and 12 qubits, respectively. In section \ref{sct:vha} the VHA is discussed in more detail and section \ref{sct:gdo} presents the simple gradient-based optimization routine and compares the two approaches for determining the gradient. Section \ref{sct:ce} details how many additional circuits are required to determine numerical and analytical gradients. A brief overview on how noise has been implemented in our simulations is given in section \ref{sct:depol}. Finally, results are shown for the simple quantum circuit in section \ref{sct:sc} and for the one-dimensional Hubbard model in section \ref{sct:hubbard}.

\section{The Variational Hamiltonian Ansatz}
\label{sct:vha}
The VHA starts from a trial state given by
\begin{equation}
  \label{eqn:vha}
  \ket{\psi(\boldsymbol{\theta})} = \hat{U}(\boldsymbol{\theta})\ket{\psi_0},
\end{equation}
with $\boldsymbol{\theta}$ defining the parameter set, $\hat{U}(\boldsymbol{\theta})$ being a parametrized unitary operator, which is implemented by a quantum circuit and $\ket{\psi_0}$ an initial state usually chosen to be a single Slater determinant of orbitals obtained from the mean-field solution. The energy of this trial state for a system with Hamiltonian $\hat{\mathcal{H}}$ can then be written as
\begin{equation}
  E(\boldsymbol{\theta}) = \bra{\psi(\boldsymbol{\theta})}\hat{\mathcal{H}}\ket{\psi(\boldsymbol{\theta})}. \label{eqn:E_theta}
\end{equation}
Minimizing the energy with respect to the parameter set leads to the optimal energy attainable with the Ansatz and the corresponding parameter set $\boldsymbol{\theta}_{\rm{opt}}$ satisfying
\begin{equation}
  E(\boldsymbol{\theta}_{\rm{opt}}) = \min_{\boldsymbol{\theta}}E(\boldsymbol{\theta}).
\end{equation}
According to the variational principle \cite{sakurai2014,mcardle2020}, the exact ground state of the Hamiltonian $E_{\rm{exact}}$ sets a lower bound for the variationally determined energy
\begin{equation}
  E_{\rm{exact}} \leq E(\boldsymbol{\theta}_{\rm{opt}}).
\end{equation}
Starting with an initial guess, $\boldsymbol{\theta}_0$, the parameter set is updated iteratively, using the procedure described in Sec.~\ref{sct:gdo}, to approach the optimal solution.

The explicit form of the unitary operator, $\hat{U}(\boldsymbol{\theta})$, is constructed by decomposing the Hamiltonian into $N$ separate terms, as stated in Eq.~\eqref{eqn:vhaham}. The partial contributions to the Hamiltonian are Hermitian operators, so we can use them as generators for unitary rotations. Since the partial contributions do not commute in general, the order in which the unitaries are applied matters. In practice we repeat the application of the unitaries $n$ times and allow for different parameters (or rotation angles) in each repetition (Eq.~\eqref{eqn:unitary}),
\begin{subequations}
  \begin{align}
    \label{eqn:vhaham}
    \hat{\mathcal{H}} & = \sum_{\alpha=1}^N \hat{\mathcal{H}}_{\alpha}, \\
    \label{eqn:unitary}
    \hat{U}(\boldsymbol{\theta}) & = \prod_{k=1}^n\prod_{\alpha=1}^N e^{i\theta_{\alpha, k}\hat{\mathcal{H}}_{\alpha}}.
  \end{align}
\end{subequations}

In this work we investigate the 1D Hubbard model which is described by the Hamiltonian
\begin{equation}
  \begin{split}
    \hat{\mathcal{H}} &= \hat{\mathcal{T}} + \hat{\mathcal{W}} \\
    &= -t\sum_{i}\sum_{\sigma}\left(\hat{c}_{i,\sigma}^{\dagger}\hat{c}_{i+1,\sigma}
    +\hat{c}_{i+1,\sigma}^{\dagger}\hat{c}_{i,\sigma}\right) \\  
    &\quad + U\sum_{i}\left(\hat{c}_{i,\uparrow}^{\dagger}\hat{c}_{i,\uparrow} - \frac{1}{2}\right)\left(\hat{c}_{i,\downarrow}^{\dagger}\hat{c}_{i,\downarrow} - \frac{1}{2}\right),
  \label{Hhub}
  \end{split}
\end{equation}
with the first term describing nearest-neighbour hopping between site $i$ and $i+1$ (kinetic energy) with hopping amplitude $t$ and the second term describing the repulsion of two electrons on the same site with interaction strength $U$. We consider a system is at half-filling -- an average of one electron per site. In our studies we focus on the case where the kinetic energy and the interaction have the same magnitude and use $t$ as our unit of energy, i.e., $U = t = 1$. Moreover, we employ periodic boundary conditions. The Hubbard model at half filling is a prototypical model describing a Mott insulator, i.e., an insulator where the fundamental gap is due to electron-electron interaction and any (static) mean-field theory would yield a metal (unless the symmetry is artificially broken).

Quite naturally Hamiltonian~\eqref{Hhub} can be split into two contributions, i.e., the kinetic energy $\hat{\mathcal{T}}$ and the interaction energy $\hat{\mathcal{W}}$. For system sizes larger than 2 sites the hopping operator is split further into a so-called ``even'' and ``odd'' contribution, $\hat{\mathcal{T}}_e$ and $\hat{\mathcal{T}}_o$. The individual contributions to each of the three parts of the Hamiltonian, $\hat{\mathcal{T}}_e$, $\hat{\mathcal{T}}_o$ and $\hat{\mathcal{W}}$, commute among each other, which implies that each exponential can be split and reordered easily without using the Baker-Campbell-Hausdorff formula.

\section{Gradient descent optimizer}
\label{sct:gdo}
In this work we are interested in investigating how noise affects the synthesis of gradients of the cost function (cf.\ Eq.\ \eqref{eqn:E_theta}). While there are several established algorithms, such as the BFGS algorithm \cite{broyden1970,fletcher1970,goldfarb1970,shanno1970}, for gradient-based optimization, we decided to use a very simple approach, namely a (damped) steepest-descent optimizer, in order to reduce the complexity in the optimization process and focus on the gradient. Specifically, we update a parameter $\theta_i$ according to
\begin{equation}
\theta^{t+1}_i = \theta^t_i - \eta\partial_{\theta_i}E(\boldsymbol{\theta}^t), \label{eqn:theta_update}
\end{equation}
where $\partial_{\theta_i}E(\boldsymbol{\theta})$ denotes the derivative of the cost function with respect to $\theta_i$, $\eta$ is a \emph{fixed} learning rate (damping) controlling the step size towards the minimum of the cost function and $\theta^t_i$ is the $i^{th}$ parameter at iteration step $t$. The parameter set at iteration $t+1$ is given by the parameter set at step $t$ modified by the cost function's gradient scaled by the learning rate $\eta$. Compared to other gradient-based optimizers\cite{nocedal2006}, the steepest-descent optimizer is very simplistic in its form, since it contains only one fixed hyperparameter. Since no information beyond the gradient, such as an approximate Hessian matrix, is used, algorithm~\eqref{eqn:theta_update} may converge slower than, for example, the aforementioned BFGS optimizer.

In order to determine the gradients for the optimization routine, two possibilities will be discussed throughout this work. On the one hand, gradients can be determined numerically with a finite-difference method:
\begin{equation}
  \begin{split}
	\partial_{\theta_i}E(\boldsymbol{\theta}) \approx \frac{1}{\epsilon} \Big(E(\theta_1,\ldots,\theta_i + \epsilon,\ldots,\theta_n) \\
    - E(\theta_1,\ldots,\theta_i,\ldots,\theta_n) \Big), \label{eqn:finitDiff}
  \end{split}
\end{equation}
with $\epsilon$ defining a small \emph{but finite} step by which the parameter $\theta_i$ is shifted. This immediately implies that the number of additional circuit evaluations for obtaining the gradient using the finite-difference method corresponds to the number of parameters. However, this method may be more susceptible towards noise effects, such as statistical or depolarization noise due to the fact the energy difference (cf.\ Eq.\ \eqref{eqn:finitDiff}) is getting smaller, and therefore harder to resolve, if we improve the accuracy of the gradient by making the step width $\epsilon$ smaller.

On the other hand we determine gradients analytically with the so-called parameter-shift rule \cite{schuld2019}. At its core the parameter-shift rule uses the fact that the cost function generally is represented by a quantum circuit composed of (parametrized) single-qubit rotations and \emph{fixed} two-qubit gates. Focusing on the dependence of the energy on a single parameter $\theta_1$, which we assume to only control \emph{one} single-qubit gate, the energy, Eq.\ \eqref{eqn:E_theta}, can be written as,
\begin{equation}
  \begin{split}
	& E(\theta_1; \theta_2,\ldots,\theta_n) = 
	A_1 \cos(\omega\theta_1+\varphi_1) + \ldots \\
	& \quad \ldots + A_\alpha \cos(\omega\theta_1+\varphi_\alpha) + \ldots + C,
	\label{eqn:energyTrig}
  \end{split}
\end{equation}
with amplitudes $A_\alpha$, phase shifts $\varphi_\alpha$ and constant $C$, which depend on all other parameters $\theta_2, \ldots, \theta_n$ (see appendix \ref{APP} for a detailed derivation). The distance between the two eigenvalues of the generator for the single-qubit rotation is denoted by $\omega$ -- for Pauli matrices we have $\omega = 2$. The main idea of the parameter-shift rule is to consider the derivative of a trigonometric function,
\begin{equation}
  \begin{split}
    \partial_{\theta}\cos(\omega\theta + \varphi) &= -\omega\sin(\omega\theta + \varphi) \\
    &= \frac{\omega}{2}\left[\cos\left(\omega\left[\theta + \frac{\pi}{2\omega}\right] + \varphi\right) \right. \\ &\quad \left. - \cos\left(\omega\left[\theta - \frac{\pi}{2\omega}\right] + \varphi\right)\right]~,
    \label{eqn:deriv}
  \end{split}
\end{equation}
which highlights that the \emph{exact} derivative with respect to the parameter $\theta$ is obtained by taking the difference of two quantum circuits, where the parameter is shifted by $\pm\frac{\pi}{2\omega}$, respectively. The parameter-shift rule, as presented above, is based on the assumption that a parameter $\theta_i$ \emph{only} controls a single single-qubit gate (cf.\ App. \eqref{APP}). However, the parameter $\theta_i$ can appear in more than one gate for the VHA Ansatz. This is due to the fact that the parameters are defined in Eq.\ \eqref{eqn:unitary} in reference to generators from the electronic Hamiltonian. In transforming the electronic Hamiltonian into the quantum circuit several single-qubit rotations are parametrized, in general, by the same parameter $\theta_i$. Hence the parameter-shift rule \emph{cannot} be simply applied to the parameters defined in Eq.\ \eqref{eqn:unitary}. Use of the parameter-shift rule can be vindicated by defining a new set of parameters, $\boldsymbol{\mu}$, which assigns each parametrized single-qubit gate its own parameter $\mu_i$. Then we can compute the gradient with respect to the new set of parameters as
\begin{equation}
  \begin{split}
    \partial_{\mu_i} E(\boldsymbol{\mu}) = r \left[E\left(\mu_1,\ldots,\mu_i+\frac{\pi}{4r},\ldots,\mu_n\right) \right. \\
    \left. - E\left(\mu_1,\ldots,\mu_i-\frac{\pi}{4r},\ldots,\mu_n\right)\right]~.
    \label{eqn:paramShiftOnce}
  \end{split}
\end{equation}
Note that the parameter $r$, controlling the ``step width'', in principle depends on the explicit form of the unitary operator, Eq.~\eqref{eqn:unitary}, but we can always define a linear map from the parameter set $\boldsymbol{\theta}$ to $\boldsymbol{\mu}$, such that $r$ is the same for all parametrized gates ($r=\frac{1}{2}$ for standard Pauli rotations, see e.g.\ Ref.\ \citen{schuld2019} for details).

We emphasize that the parameter-shift rule yields, in principle, the exact gradient and not an approximation as the finite-difference method described above. Moreover, the parameter-shift rule is potentially more resilient towards noise and other effects, because -- in spite of being exact -- it is evaluated using a finite difference with a step width on the order of the spectral width of the generator (typically a scaled Pauli matrix). However, the number of circuit evaluations scales with the number of parametrized gates and thus, with increasing system size, requires more circuits than the finite-difference method. In our work we consider a specific Ansatz, the VHA, for generating the variational quantum circuit, which requires discussing the difference between the number of parameters (components of $\boldsymbol{\theta}$) and the number of parametrized gates (components of $\boldsymbol{\mu}$). For a generic variational quantum circuit, which aims at optimizing all single-qubit gates independently, the number of parameters coincides trivially with the number of parametrized gates.

\section{Circuit evaluation}
\label{sct:ce}
In this section, we highlight the difference between the number of parametrized gates and parameters mentioned in the previous section to expose the additional overhead for using the parameter-shift rule. An overview for the one-dimensional Hubbard model is given in Table \ref{tab1}, where we compare the number of parameters and the number of parametrized gates for 1D Hubbard rings of various sizes (number of sites). Therefore a circuit was generated using a two-qubit decomposition \cite{Bauer2016, Vatan2004}, consisting of one-qubit rotation gates and a two-qubit controlled-Z gate.
\begin{center}
  \begin{table}
    \begin{tabular}{c c c c}
	  \hline
      \hline
      $\#$ sites $M$ & \thead{Hilbert space size \\ $2^{2M}$} & $\#$ parameters & \thead{$\#$ parametrized \\ gates} \\
      \hline    
      2 & 16 & 2 & 10 \\
      4 & 256 & 3 & 28 \\
      6 & 4,096 & 3 & 42 \\
      8 & 65,536 & 3 & 56 \\
      10 & 1,048,576 & 3 & 70 \\
      12 & 16,777,216 & 3 & 84 \\
      14 & 268,435,456 & 3 & 98 \\
      16 & 4,294,967,296 & 3 & 112 \\
      \hline
      \hline
    \end{tabular}
    \caption{\label{tab1} Number of parameters and parametrized gates for 1D Hubbard chains with varying number of sites.}
  \end{table}
\end{center}
Table \ref{tab1} shows the number of parameters and parametrized circuits for various sizes, determined by the number of sites $M$ in the one-dimensional Hubbard model. In the case of $M=2$ with one repetition for the Ansatz made in Eq.~\eqref{eqn:unitary}, we only have two parameters, namely one parameter for the hopping operator and one for the interaction operator. Increasing the site number to $M=4$ or $M=6$ yields one further parameter per repetition, since we now also split the hopping term (kinetic energy) into two internally commuting contributions (labelled ``even'' and ``odd''). The total number of circuits to evaluate for calculating energy and gradient with the finite-difference method is given by the number of parameters, i.e., one circuit for each parameter shifted by $\epsilon$, plus the circuit with no shift applied to any parameter. Considering the 2-site Hubbard model with one repetition in the VHA, which implies two parameters, a total number of three circuits would be required in order to evaluate the gradient and the energy. Increasing the number of repetitions from one to two, leading to four parameters, 5 circuits have to be evaluated in order to determine the energy and its gradient. In general, the total number of circuit evaluations required for gradient and energy determination is given by $N_{\rm{fd}} = RP + 1$ when using the finite-difference method, where $P$ defines the number of parameters and $R$ the number of repetitions in the VHA.

For the parameter-shift rule, the additional circuits for evaluating the gradient is proportional to twice the number of parametrized gates, leading to 21 circuits for the 2-site Hubbard model with one repetition. Similarly to the finite-difference case, the number of parametrized gates scales linearly with the number of parameters and therefore also with the number of repetitions. In general, the number of circuits required using the parameter-shift rule can be written as $N_{\rm{ps}} = 2C(M)RP + 1$, where we introduced the coefficient $C(M)$, which translates the number of parameters to the number of parametrized gates. Considering the data in table \ref{tab1} and system sizes larger than 2 sites, we get $C(M) = \frac{7}{3}M$, which implies that the number of additional circuits scales linearly with the system size using the Jordan-Wigner transformation \cite{Jordan1928} to map fermionic sites to qubits.

\section{Depolarization noise}
\label{sct:depol}
\begin{figure}[h!]
  \centering
  \includegraphics[clip,width=8.6cm]{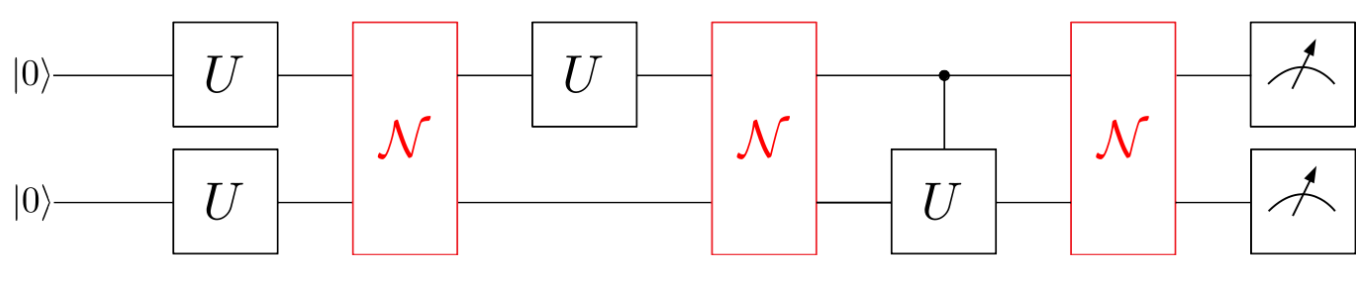}
  \caption{Graphical illustration of a simple two-qubit circuit with one- and two-qubit unitary operations $U$ and three noise gates, labeled by $\mathcal{N}$, describing the effect of depolarization.}
  \label{fgr:circuit}
\end{figure}

In this section we describe how we simulate depolarization noise in the quantum device. Since a quantum device can effectively be considered as an open quantum system, i.e., the qubits on the quantum chip interacting with the environment, we simulate the execution of the circuits as follows: the quantum circuit is arranged in such a way, that as many gates as possible are executed in parallel. After that the noise gate is applied to all qubits in the circuit, which can effectively be described as the application of Kraus operators to all qubits. This procedure is repeated until the measurement of observables is performed and is illustrated in Fig.~\ref{fgr:circuit}.

The noise channel $\mathcal{\hat{N}}$, acting on a single qubit, can be described as \cite{king2003}
\begin{equation}
\mathcal{\hat{N}}(\hat{\rho}) = \sum_{i=0}^3 \hat{K}_i \hat{\rho} \hat{K}_i^{\dagger},
\end{equation}
with $\hat{\rho}$ being the density matrix representing the quantum system. The Kraus operators $\hat{K}_i$ are defined as follows:
\begin{subequations}
	\begin{align}
	\label{eqn:k0}
	\hat{K}_0 = \sqrt{1-\frac{3}{4}\Gamma}\hat{I}, \\
	\label{eqn:ki}
	\hat{K}_i = \frac{\sqrt{\Gamma}}{2}\hat{\sigma}_i, i \in \{1,2,3\},
	\end{align}
\end{subequations}
with $\Gamma = 1 - e^{-\gamma}$ defining the damping term and $\hat{\sigma}_i$ being the Pauli operators. It can be easily seen that in the noiseless case, i.e. $\gamma = 0$, the damping term $\Gamma$ vanishes and only the Kraus operator with the identity operator $\hat{I}$ is left, leading to a noiseless quantum operation $\mathcal{\hat{N}}(\hat{\rho}) = \hat{\rho}$. In the noisy case with $\gamma > 0$, not only $\hat{K}_0$, but also the three other Kraus operators $\hat{K}_i$ including the Pauli operators are taken into account, so that the quantum operation $\mathcal{\hat{N}}$ induces depolarization in the system. The parameter $\gamma$, determining the strength of the noise, is estimated as the ratio of the gate time $T_g$ and the coherence time $T_2$ of a qubit.

The gate time $T_g$ represents the time it takes to perform a given operation, for instance the duration of a microwave pulse for controlling superconducting qubits. In Fig.\ \ref{fgr:circuit} the gate time can be viewed as the width of a single gate column, if we interpret the horizontal direction of the algorithm as time axis. An example for a gate operation is a qubit flip, where a qubit prepared in the $\ket{0}$ state is flipped to the $\ket{1}$ state after applying an X gate on that qubit. For superconducting architectures, usual single and two-qubit gate times range from $5 - \SI{500}{ns}$ \cite{devoret2013,schutjens2013}, whereas trapped-ion designs show single-qubits gate times in the microseconds and two-qubit gate times in the $10 - \SI{100}{\mu s}$ regime \cite{bruzewicz2019,linke2017}. The depolarization time $T_1$ determines the time it takes for a \emph{undesired} qubit flip to occur, due to the coupling of the qubit to its environment. Hence, if we would like to perform $N$ operations on a qubit, we require $NT_g \ll T_1$ or equivalently $\gamma \ll \frac{1}{N}$. In passing we note that the dephasing time $T_2$ is limited by the relaxation time $T_1$ of the qubit, i.e., $T_2 \leq 2 \cdot T_1$. While typical coherence times have values in the order of $\SI{100}{\mu s}$ for superconducting qubits \cite{devoret2013,rigetti2012,novikov2013,nguyen2019}, these values are several orders of magnitude higher for ion-trapped devices, which can be several seconds \cite{bruzewicz2019}. Considering these numbers, the ratio $\gamma$, which characterizes the noise strength, in our simulations is chosen within
\begin{equation}
  \label{eqn:gamma}
  \gamma = \frac{T_g}{T_2} \in \big[ 10^{-4}, 10^{-2} \big].
\end{equation}

\section{Numerical Results}
\subsection{Simple circuit}
\label{sct:sc}
In order to get a better understanding of the gradient descent optimizer, a simple one-qubit circuit has been investigated, where at first a Hadamard gate is applied to the prepared ground state $\ket{0}$ followed by a rotation Z gate (see Eq.\ \eqref{eqn:cosa}). After this operation, a measurement of the Pauli X operator is performed, leading to a trivial periodic function
\begin{subequations}
\begin{align}
\label{eqn:cosa} 
\begin{split}
\ket{\psi(\theta)} &= \hat{U}(\theta)\ket{\psi_0} = \hat{R}_z(\theta)\hat{H}\ket{\psi_0} \\
&=
\begin{pmatrix}
	e^{-i\theta/2} & 0 \\
	0 & e^{i\theta/2}
\end{pmatrix}\frac{1}{\sqrt{2}}\begin{pmatrix}
	1 & 1 \\
	1 & -1
\end{pmatrix} \begin{pmatrix}
	1 \\
	0
\end{pmatrix} 
\\
&= \frac{1}{\sqrt{2}} \begin{pmatrix}
	e^{-i\theta/2} \\
	e^{i\theta/2}
\end{pmatrix},
\end{split} \\
\label{eqn:cos}
E(\theta) &= \braket{\hat{X}} = \bra{\psi(\theta)}\hat{X}\ket{\psi(\theta)} = \cos(\theta).
\end{align}
\end{subequations}

The minimum of Eq.~\eqref{eqn:cos} is attained for $\theta = (1 + 2n) \pi$ ($ n \in \mathbb{Z})$ with a minimal value of $E_{\rm{exact}} = -1$.

\subsubsection*{Shot noise}
\begin{figure}[h!]
  \centering
    \includegraphics[clip,width=8.6cm]{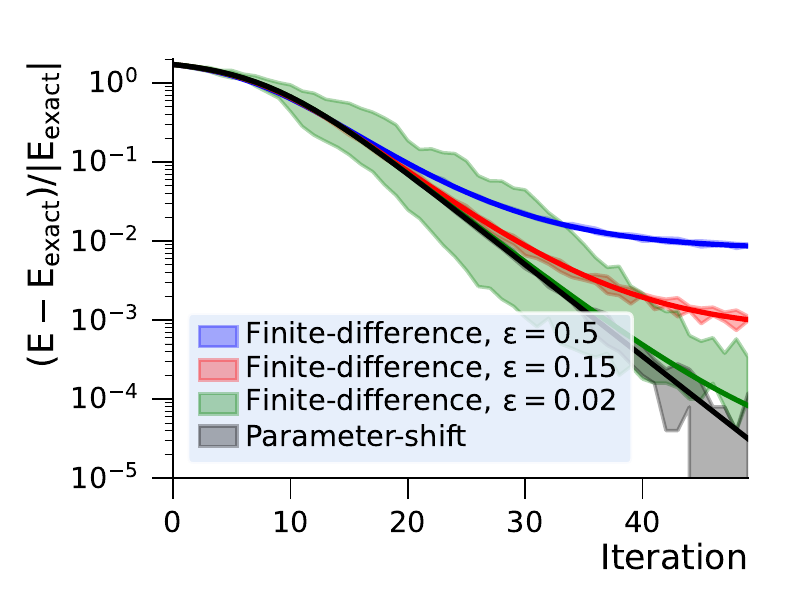}
    \caption{Optimization runs using a simple gradient descent Ansatz performed with the finite-difference method compared to the parameter-shift rule for the simple circuit and $N=50000$. The bold line indicates a run without shot noise, whereas the corresponding shaded region marks five optimization runs performed with shot noise. For all optimization runs, a learning rate of $\eta=0.5$ has been used.}
    \label{fgr:simple}
\end{figure}

Figure \ref{fgr:simple} shows optimization runs for different values of $\epsilon$ for the finite-difference method, as well as runs performed with the parameter-shift rule with $N=50000$ measurement shots. Note that in all plots, starting from Fig.~\ref{fgr:simple}, we are using a logarithmic scale for the energy differences shown on the y-axis. The solid lines indicate simulations without any source of noise ($N \to \infty$) and the shaded areas with the corresponding color are obtained as the envelope of five runs performed with statistical errors ($N = 50000$), i.e., the upper line marks the maximum and the bottom line indicates the minimum of the relative deviation from the optimal value at a given iteration of the optimization, respectively. Focusing on the $N\to\infty$ results, it can be seen that with decreasing $\epsilon$ the accuracy gets better, which highlights the error introduced by approximating the derivative using a finite difference. However, if statistical effects are included it can also be seen that the fluctuations around the optimal curve increase with smaller $\epsilon$, clearly exposing that smaller $\epsilon$ are more prone to statistical errors. Considering the optimization runs performed with the parameter-shift rule it can be seen that the on the one hand, a more accurate result is achieved for the noiseless run ($N \to \infty$), demonstrating that the parameter-shift rule yields the exact gradient, and on the other hand the fluctuations around the optimal curve are smaller compared to the finite-difference runs with $\epsilon=0.02$.

\subsubsection*{Shot and depolarization noise}
\begin{figure}[h!]
  \centering
    \includegraphics[clip,width=8.6cm]{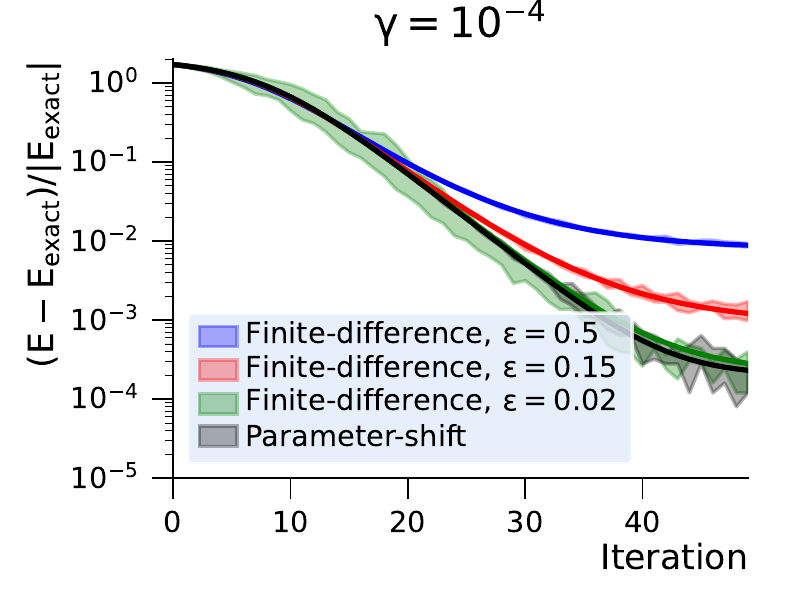}
    \caption{Same plot as Fig.\ \ref{fgr:simple}, but with depolarization noise, characterized by the rate $\gamma = 10^{-4}$ included in the simulation.}
    \label{fgr:simple_noise_small}
\end{figure}

\begin{figure}[h!]
  \centering
    \includegraphics[clip,width=8.6cm]{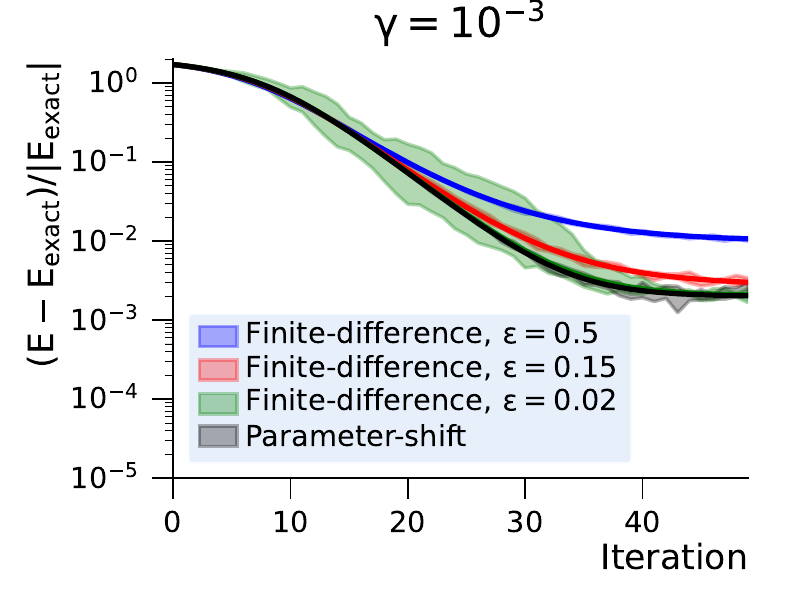}
    \caption{Same plot as Fig.\ \ref{fgr:simple_noise_small}, but with a depolarization rate of $\gamma = 10^{-3}$.}
    \label{fgr:simple_noise_medium}
\end{figure}

\begin{figure}[h!]
  \centering
    \includegraphics[clip,width=8.6cm]{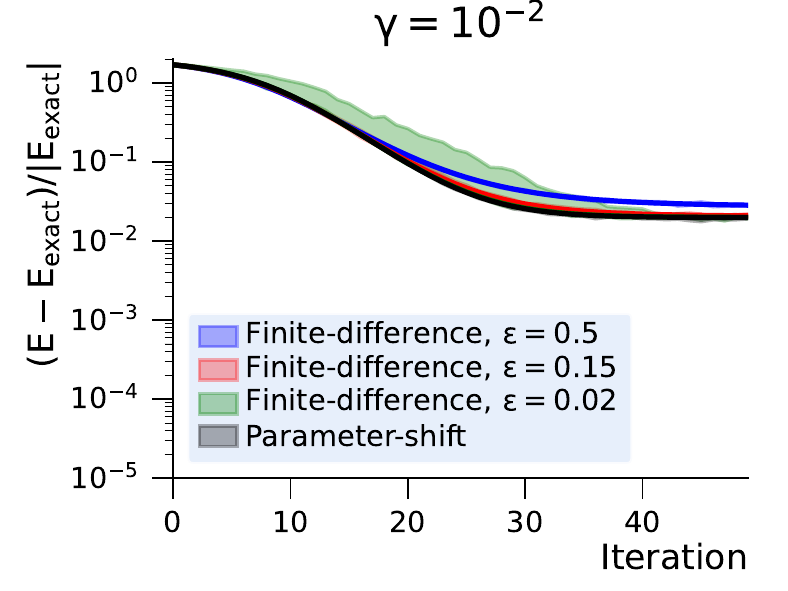}
    \caption{Same plot as Fig.\ \ref{fgr:simple_noise_small}, but with a depolarization rate of $\gamma = 10^{-2}$.}
    \label{fgr:simple_noise_large}
\end{figure}

Including the effect of depolarization noise, we observe that the deviation from the cost function's minimum increases. This is most pronounced for the parameter-shift rule and small values of $\epsilon$ for the finite-difference method. We can attribute this to the fact that the error introduced by using a larger $\epsilon$ in the finite difference dominates the overall error. This is depicted in Fig.~\ref{fgr:simple_noise_small}, where for the two mentioned cases a small shift away from the minimum can be seen, compared to the noiseless run in Fig.~\ref{fgr:simple}. Increasing the depolarization rate further leads to a higher deviation from the minimum for all cases, as can be seen in Figs.~\ref{fgr:simple_noise_medium} and \ref{fgr:simple_noise_large}. Especially for the case where $\gamma=10^{-2}$, the finite-difference curves with the two smallest $\epsilon$ and the parameter-shift curve nearly show the same result for the cost function after 50 iteration steps, clearly showing that the depolarization error dominates the overall error.

\subsection{The 1D Hubbard model}
\label{sct:hubbard}
Next, we study the one-dimensional Hubbard model by performing optimization runs with shot noise only and with shot and depolarization noise. The Hubbard Hamiltonian is defined in Eq.~\eqref{Hhub}, where the initial state, $| \psi_0 \rangle$, is chosen as the ground-state Slater determinant for the non-interacting system.

\subsection*{Results for 2 sites}
For the 2-site Hubbard model we decompose the Hamiltonian simply into kinetic energy and the interaction energy. Thus, the trial state for the 2-site Hubbard model is given as
\begin{equation}
  \label{eqn:2site}
  \ket{\psi(\boldsymbol{\theta})} = e^{i\theta_2\hat{\mathcal{T}}} e^{i\theta_1\hat{\mathcal{W}}}\ket{\psi_0},
\end{equation}
where we only use one repetition of the unitaries, since this parametrization already encompasses the exact ground state for the 2-site Hubbard model. As in the previous example the parameter set $\boldsymbol{\theta}$ has been determined with the gradient descent method explained in section~\ref{sct:gdo}. The exact analytical result for the 2-site Hubbard model \cite{matlak2003} with Eq.~\eqref{eqn:2site} as the trial state has been taken as the reference value $E_{\rm{exact}}$.

\subsubsection*{Shot noise}
\begin{figure}[h!]
  \centering
    \includegraphics[clip,width=8.6cm]{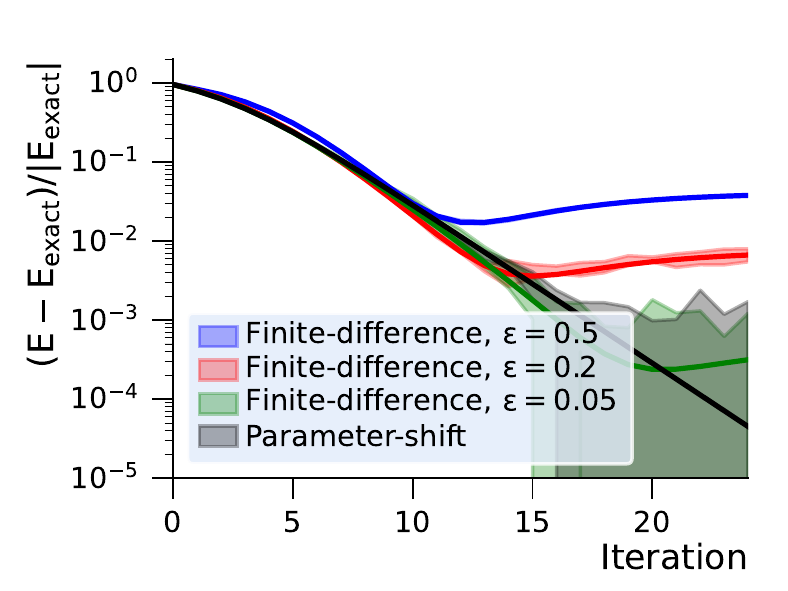}
    \caption{Optimization runs using a simple gradient descent Ansatz performed with the finite-difference method compared to the parameter-shift rule for the 2-site Hubbard model and $N=50000$. The solid lines indicate runs without shot noise, whereas the corresponding shaded region marks five optimization runs performed with shot noise. For all optimization runs, a learning rate of $\eta=0.1$ has been used.}
    \label{fgr:2site}
\end{figure}

At first, optimization runs including only shot noise have been performed, where the number of measurement shots has been set to $N=50000$. As in the previous case the solid lines in Fig.~\ref{fgr:2site} show runs without shot noise, corresponding to the limit $N\rightarrow\infty$. The shaded areas represent the maximum and minimum of five optimization runs performed with shot noise, respectively. From the graph it can be seen that the shaded region for $\epsilon=0.5$ is vanishingly small at the scale of the plot. Starting around iteration step 10 it can be observed, at the scale of the plot, that the fluctuations of the shaded regions increase with smaller $\epsilon$. The more accurate the gradients are in principle, the more relevant becomes the fact that in practice we always perform a finite amount of measurements. Fig.~\ref{fgr:2site} shows that the energy deviation exhibits a minimum for gradients obtained via finite difference, which can be attributed to the fact that the approximate derivatives do not even in principle correspond to the true derivatives of the cost function. This has to be contrasted with the energy deviation obtained by using the parameter-shift rule to synthesize the gradients for the optimization procedure. Here, the gradients are in principle ($N \to \infty$) exact and the only error stems from the finite accuracy due to a finite number of measurements for the gradient. However, our results show that for a fixed number of measurements ($N=50000$) the intrinsic error in calculating the gradients via finite difference using $\epsilon=0.05$ is smaller than the statistical error due to the finite number of measurements. Considering the measurement overhead (cf.\ Tab.\ \ref{tab1} and Sec.\ \ref{sct:ce}) in synthesizing the gradients via the parameter-shift rule, favors the finite-difference gradients.

\subsubsection*{Shot and depolarization noise}
\begin{figure}[h!]
  \centering
    \includegraphics[clip,width=8.6cm]{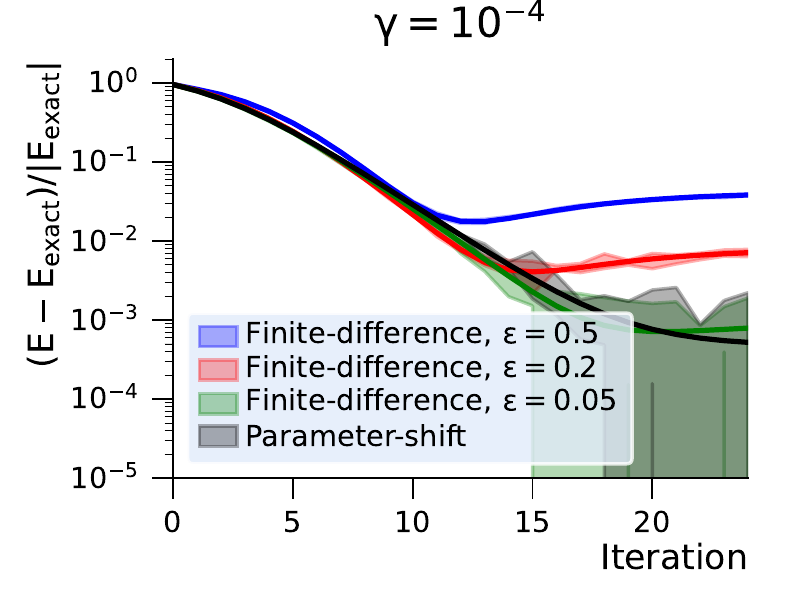}
    \caption{Same as Fig.\ \ref{fgr:2site}, but now including depolarization noise, characterized by the rate $\gamma = 10^{-4}$, in the simulations.}
    \label{fgr:2site_noise}
\end{figure}

By adding depolarization noise with a rate of $\gamma=10^{-4}$ to the previous simulations it can be seen that there are no significant differences for $\epsilon=0.5$ and $\epsilon=0.2$, except a minor increase in the deviation from the exact energy, as shown in Fig.~\ref{fgr:2site_noise}. However, there is a larger offset for the parameter-shift run. For example, around the 20th iteration step, the $N \to \infty$ parameter-shift curve flattens out. As in the previous case, parameter-shift simulations show a similar behaviour as the finite-difference runs with small $\epsilon$.

\subsection*{Results for 6 sites}
In contrast to the 2-site Hubbard model, the number of repetitions has been set to 2 for the 6-site Hubbard model, leading to a trial state given as
\begin{equation}
  \label{eqn:6site}
  \ket{\psi(\boldsymbol{\theta})} = e^{i\theta_6\hat{\mathcal{T}}_o} e^{i\theta_5\hat{\mathcal{T}}_e} e^{i\theta_4\hat{\mathcal{W}}} e^{i\theta_3\hat{\mathcal{T}}_o} e^{i\theta_2\hat{\mathcal{T}}_e} e^{i\theta_1\hat{\mathcal{W}}}\ket{\psi_0},
\end{equation}
with $\hat{\mathcal{T}}_e$ and $\hat{\mathcal{T}}_o$ defined in Sec.\ \ref{sct:vha}.

By increasing the number of repetitions and thus the number of parameters in the Ansatz, the variationally determined energy is closer to the true ground state energy of the system. Note that we use the optimal energy achievable with the given Ansatz ~\eqref{eqn:6site} as reference energy and denote it as $E_{\rm{exact}}$. Due to the system size, runs with depolarization noise were not performed.

\subsubsection*{Shot noise}
\begin{figure}[h!]
  \centering
    \includegraphics[clip,width=8.6cm]{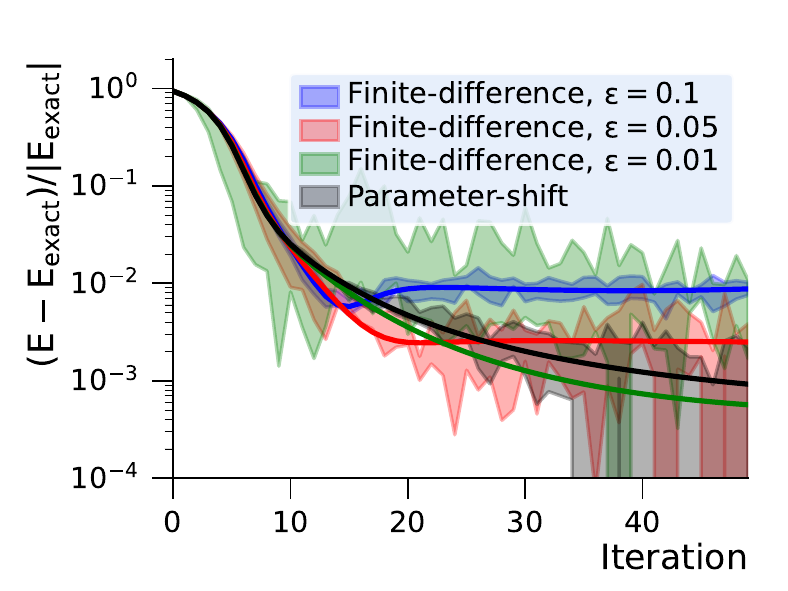}
    \caption{Optimization runs using a simple gradient descent Ansatz performed with the finite-difference method compared to the parameter-shift rule for the 6-site Hubbard model and $N=50000$. The solid lines indicate runs without shot noise, whereas the corresponding shaded region marks five optimization runs performed with shot noise. For all optimization runs, a learning rate of $\eta=0.03$ has been used.}
    \label{fgr:6site}
\end{figure}

For the  6-site Hubbard model we perform optimization runs for three different choices of $\epsilon$ and the parameter-shift rule with $N=50000$. Taking a look at the solid lines in Fig.~\ref{fgr:6site}, representing noiseless runs, it can be seen that with decreasing $\epsilon$ the energy accuracy improves, which could also be observed in the two previous cases. Taking shot noise into account, some minor fluctuations around the optimal curve for $\epsilon=0.1$ can be spotted. However, these fluctuations increase with decreasing $\epsilon$ besides that an offset seems to occur, especially for $\epsilon=0.01$. While the noiseless run suggests an energy accuracy in the order of $10^{-3}$ after 50 iterations steps with the gradient-descent optimizer, this value roughly increases by an order of magnitude for the noisy runs.

The noiseless parameter-shift run shows a worse accuracy in energy after the first 50 iterations, however it shows better results compared to the finite-difference method when shot noise is included. Fluctuations occur only around the noiseless run, whereas no offset, as in the case of $\epsilon=0.01$ performed with the finite-difference method, can be observed and thus showing an overall better performance compared to its numerical counterpart. It has to be emphasized that the total number of circuits is considerably higher for the parameter-shift rule -- namely a factor of 24 -- for this particular system and consequently requires a lot more time to finish the 50 iterations compared to the finite-difference method. Therefore, a compromise between runtime and accuracy in energy is a plausible conclusion. Having a slightly worse accuracy but therefore requiring a smaller number of additional circuits, one might choose the finite-difference method with $\epsilon=0.05$.

\section{Conclusion}
In this work, two possibilities for gradient determination have been investigated and tested on a simple toy model and the one-dimensional Hubbard model using the Variational Hamiltonian Ansatz and a simple gradient-descent algorithm for parameter optimization.

On the one hand, gradients have been determined numerically with a simple finite-difference method, where the difference between the obtained and exact minimum of the cost function for noiseless runs gets smaller with decreasing step sizes $\epsilon$. The number of circuit evaluations scales only with the number of parameters defined in the parameter set and thus shows a fast runtime compared to the parameter-shift method.

On the other hand, analytical gradients, determined with the parameter-shift rule, lead to more accurate results and a more resilient behavior towards statistical noise and depolarizing effects. Especially in the case of the 6-site Hubbard model, a better accuracy could be achieved when simulating the system with shot noise. A major drawback of the parameter-shift rule, applied to the optimization of a VHA circuit, is its scaling. Since the VHA does not only scale with the number of parameters but also with the system size (which affects the number of parametrized gates), the runtime increases fast. For example, in the 6-site case the number of circuit evaluations is almost two orders of magnitude higher for the parameter-shift method compared to the numerical counterpart. Thus, a compromise between runtime and accuracy can be made for the finite-difference method, where the step size $\epsilon$ should be chosen optimally to avoid statistical noise and provide the required accuracy. The obvious downside of this approach that this optimal step width has to be determined and in general depends on the number of measurements and other hyper-parameters of the optimization algorithm. Taking a look at the simple toy model and the 2-site Hubbard model it can be concluded that the finite-difference method is the method of choice when weighing both runtime and accuracy. Even with shot and depolarization noise for both cases, there is no substantial difference between optimization runs performed with the finite-difference method choosing small step sizes $\epsilon$ and the parameter-shift rule.

We emphasize that the results presented in this work apply to the generic situation where the trial state is constructed with the goal of reducing the number of parameters optimized in the classical optimization loop of a quantum-classical hybrid optimization procedure, which leads to the important difference of number of parameters and number of parametrized gates when analyzing the measurement costs when considering the parameter-shift rule. One motivation for reducing the number of parameters in the trial state is given by the fact that so-called Barren plateaus \cite{mcclean2018,grant2019} hamper the optimization of functions defined in high-dimensional spaces using gradient-based algorithms. However, the trigonometric building blocks of any quantum circuit, highlighted in the discussion of the parameter-shift rule in Sec.~\ref{sct:gdo}, suggests to use alternative (gradient-free) optimization algorithms \cite{ostaszewski2021}.

Finally, another possibility is to generate the trial state using the VHA, but optimizing each parametrized gate individually in the optimization procedure. An interesting question for future studies is to investigate whether treating each parametrized gate independently considerably improves the minimal achievable energy compared to the standard VHA.

\emph{Acknowledgements}
This work was supported by the Federal Ministry of Economic Affairs and Energy, Germany, as part of the PlanQK project (01MK20005H) developing a platform and ecosystem for quantum-assisted artificial intelligence (see planqk.de).

\appendix
\section{Derivation of the parameter-shift rule}\label{APP}
In this appendix we present an explicit derivation of Eq.\ \eqref{eqn:energyTrig}. We start by choosing a specific one-qubit rotation, $\hat{R}_{\theta_i}$, in order to decompose the parametrized unitary quantum circuit into
\begin{align}
  \hat{U}(\boldsymbol{\theta}) = \hat{V} \hat{R}_{\theta_i} \hat{W} ~, \label{Udecomposition}
\end{align}
where $\hat{V}$ and $\hat{W}$ are unitary rotations parametrized by all angles $\theta_j$ except for the angle $\theta_i$, which \emph{only} parametrizes the selected single-qubit gate. The single-qubit rotation, $\hat{R}_{\theta_i}$, is explicitly given by
\begin{align}
  \hat{R}_{\theta_i} = e^{-i \theta_i \boldsymbol{n} \cdot \boldsymbol{\sigma}}
  = \cos(\theta_i) \mathrm{\bf  1} - i \sin(\theta_i) \boldsymbol{n} \cdot \boldsymbol{\sigma} ~, \label{SU2rot}
\end{align}
where $\boldsymbol{n}$ is a unit vector defining the rotation axis (e.g., Cartesian unit vectors would result in rotations around the x-, y-, or z-axis), $\theta_i$ provides the rotation angle and $\boldsymbol{\sigma}$ denotes the vector of elementary Pauli matrices. Note that the Pauli matrices are, in principle, also labeled by the qubit index on which they act, which here is suppress in order to keep the notation concise.

The cost function ~\eqref{eqn:E_theta} can be rewritten as
\begin{align}
   E(\boldsymbol{\theta}) & = \sum_m \bra{\psi_0} \hat{U}(\boldsymbol{\theta})^\dagger \hat{O}_m \hat{U} (\boldsymbol{\theta}) \ket{\psi_0} ~, \label{H_sum}
\end{align}
highlighting that the energy is synthesized by measuring several Hermitian operators $\hat{O}_m$. Defining $\ket{\Psi} = \hat{W} \ket{\psi_0}$ and $\hat{O}^\prime_m = \hat{V}^\dagger \hat{O}_m \hat{V}$ a single term of the sum ~\eqref{H_sum}
is given by
\begin{align}
  E_m = \bra{\Psi} \hat{R}^\dagger_{\theta_i} \hat{O}^\prime_m \hat{R}_{\theta_i} \ket{\Psi} ~.
\end{align}
Using Eq.\ \eqref{SU2rot} this leads to
\begin{align}
\begin{split}
E_m & = \bra{\Psi} \big[ \cos(\theta_i) \mathrm{\bf  1} + i \sin(\theta_i) \boldsymbol{n} \cdot \boldsymbol{\sigma} \big] \hat{O}^\prime_m \\
&\quad \big[ \cos(\theta_i) \mathrm{\bf  1} - i \sin(\theta_i) \boldsymbol{n} \cdot \boldsymbol{\sigma} \big] \ket{\Psi} \nonumber \\
& = \cos^2(\theta_i) \bra{\Psi} \hat{O}^\prime_m \ket{\Psi} \\
&\quad + \sin(\theta_i) \cos(\theta_i)
\bra{\Psi} i \Big[\boldsymbol{n} \cdot \boldsymbol{\sigma}, \hat{O}^\prime_m \Big] \ket{\Psi} \nonumber \\
&\quad + \sin^2(\theta_i) \bra{\Psi} \boldsymbol{n} \cdot \boldsymbol{\sigma} \hat{O}^\prime_m
\boldsymbol{n} \cdot \boldsymbol{\sigma} \ket{\Psi} ~,
\end{split}
\end{align}
where all three expectation values are given in terms of Hermitian operators. By virtue of standard trigonometric identities this can be expressed as
\begin{align}
  E_m = A_m \cos(2 \theta_i + \varphi_m) + C_m ~. \label{Em}
\end{align}
We stress that $A_m$, $\varphi_m$ and $C_m$ depend on all angles $\theta_j$ with $j \neq i$, since our initial assumption stated that $\theta_i$ does only appear in the explicitly selected single-qubit gate.
The generalization to Pauli matrices scaled by a factor $s$ can be obtained by replacing $\theta_i \to s \theta_i$. Defining $\omega = 2 s$ and summing all constant contributions $C_m$ leads to Eq.\ \eqref{eqn:energyTrig}.

\bibliography{vha_optimization.bib}

\end{document}